# A MODAL SERIES REPRESENTATION OF GENESIO CHAOTIC SYSTEM

H. Ramezanpour[1], B. Razeghi[2], G. Darmani[3], S. Noei[4], A. Sargolzaei[5]

[1]Tübingen University, Tübingen, Germany
`hamidreza.ramezanpour@gmail.com`
[2]Sadjad Institute of Higher Education, Mashhad, Iran
b.razeghi168@sadjad.ac.ir
[3]Tübingen University, Tübingen, Germany
`hamidreza.ramezanpour@gmail.com`
[4]Tabriz University, Tabriz, Iran
`noei.shirin@gmail.com`
[5]Florida International University, Miami, Florida
`a.sargolzaei@gmail.com`

## ABSTRACT

*In this paper an analytic approach is devised to represent, and study the behavior of, nonlinear dynamic chaotic Genesio system using general nonlinear modal representation. In this approach, the original nonlinear ordinary differential equations (ODEs) of model transforms to a sequence of linear time-invariant ODEs. By solving the proposed linear ODEs sequence, the exact solution of the original nonlinear problem is determined in terms of uniformly convergent series. Also an efficient algorithm with low computational complexity and high accuracy is presented to find the approximate solution. Simulation results indicate the effectiveness of the proposed method.*

## KEYWORDS

## 1. INTRODUCTION

Generally, it is difficult to obtain exact solutions for nonlinear ordinary differential equations. In most cases, only approximate solutions (either numerical solutions or analytical solutions) can be expected. There are many techniques available for the numerical solution of ODEs. Typical of them are various shooting and multi-shooting approaches [1-2], various versions of finite difference or collocation [3]. Methods for solving ODEs usually require users to provide an initial guess for the unknown initial states and/or parameters. A common limitation to all of these approaches is that they can, at best, achieve convergence to a local solution of the ODEs, which means other solutions of interest may be missed. In recent years, much attention has been devoted to the newly developed methods to construct an analytic solutions of nonlinear equation, such methods include Adomian decomposition method (ADM) [4,5,6] and the Variational Iteration Method (VIM) [7-8,9], the homotopy analysis method (HAM)[10,11] and the homotopy-perturbation method (HPM) which is combination of perturbation method and homotopy method [12,13,14,15]. All these method have their own limitations and advantages. Traditional perturbation methods [12] strongly depend on the existence of small/large parameters. In non-perturbation methods sometimes the series solutions obtained diverge or the convergence region of their series solution is generally small and there is no proof of convergence series which obtain by these methods. In some cases their series diverge with altering some unknown constant parameters in the problem or with expanding the time span of the problem. Initial guesses and choosing linear and nonlinear operators in HPM and HAM can be very challenging and the convergence of series solutions is very sensitive to them.

Modal series [16,17], a proposed approach in nonlinear system analysis, extends the concepts of linear system theory to gain better understanding of the nonlinear systems. This approach provides the solution of autonomous nonlinear systems in terms of the fundamental and interacting modes and yields a good deal of physical insight into the system behavior [18]. Different from the linearization method, validity and accuracy of the Modal series is not restricted to a small neighborhood of the operating point. Also, in contrast to the perturbation techniques, Modal series method is independent upon small/large physical parameters in system's model. Although the other traditional non-perturbation techniques are formally independent of small/large physical parameters, they can not ensure the convergence of solution series. Unlike all perturbation and traditional non-perturbation methods, solution obtained in the Modal series form converges uniformly to the exact solution. However, it has limitations with hard nonlinearities which do not have an analytic model.

The solution of Genesio system was considered by different researchers such as Goh et al [19], used VIM and MVIM and Bataineh et al [20] obtained the solution using homotopy analysis method. The state space model of system is as follows:

$$\begin{cases} \dot{x}(t) = y(t) \\ \dot{y}(t) = z(t) \\ \dot{z}(t) = -cx(t) - by(t) - az(t) + x^2(t) \end{cases} \quad (1)$$

subject to the initial conditions:
$x(0) = 0.2$ , $y(0) = -0.3$ , $z(0) = 1$  (2)

where $x$, $y$, $z$ are state variables and $a$, $b$, $c$, are positive constants satisfying, $ab < c$.
Bifurcation study shows that when $a = 1.2$, $b = 2.92$ and $c = 6$, the above system is chaotic.

In this study for the first time an algorithm based on Modal series is proposed to solve Genesio chaotic system and to find an approximate solution for it, analytically. The rest of the paper is organized as follows. In section 2 principle concepts of Modal series is discussed. Section 3 elaborates how to find an approximate solution through an easy handling iterative algorithm. In section 4 we present the simulation results, and finally conclusions are given in the last section.

## 2. MODAL SERIES

A wide class of nonlinear dynamic systems can be modeled by the differential equations in the form of:
$$\dot{X}(t) = G(X(t)) \quad (3)$$

where $X$ is the $N$-dimensional state vector and $G : R^N \longrightarrow R^N$ is a smooth vector field and $X_{ini}$ is the vector of initial conditions. Expansion of $G$ in the taylor series around the origin and using again $X$ and $x_i$ as the new state vector and state variable respectively, yields the following representation:

$$\dot{x}_i = A_i X + \frac{1}{2}\sum_{k=1}^{N}\sum_{l=1}^{N} H_{kl}^i x_k x_l + \frac{1}{6}\sum_{p=1}^{N}\sum_{q=1}^{N}\sum_{r=1}^{N} E_{pqr}^i x_p x_q x_r + ... \quad (4)$$

Where $x$ belongs to the convergence domain of the taylor series; $v \subseteq R^N$, $A_i = \left.\frac{\partial G}{\partial x}\right|_{x=0}$ is the $i^{th}$ row of Jacobian matrix; $H^i = \left.\frac{\partial^2 G_i}{\partial x_k \partial x_l}\right|_{x=0}$, $E^i = \left.\frac{\partial^3 G_i}{\partial x_p \partial x_q \partial x_r}\right|_{x=0}$, and so on and $G_i$ is the $i^{th}$ element of vector field $G$. Assuming the system has $N$ distinct eigenvalues, $\lambda_j$, $j = 1,2,...,N$, and denoting by $U$ and $V$ the matrices of the right and left eigenvectors of $A$, respectively, the transformation $X = UY$ yield the following equivalent system for (4):

$$\dot{y}_j(t) = \lambda_j y_j(t) + \sum_{k=1}^{N}\sum_{l=1}^{N} C_{kl}^j y_k(t) y_l(t) + \sum_{p=1}^{N}\sum_{q=1}^{N}\sum_{r=1}^{N} D_{pqr}^j y_p(t) y_q(t) y_r(t) + ... \quad (5)$$

where $Y$ belongs to the linear mapping of $v$ denoted by $v \subseteq C^N$ under defined linear transformation, here and from now on $j = 1,2,...,N$,

$$C^j = \frac{1}{2}\sum_{p=1}^{N} V_{jp}^T [U^T H^p U] = [C_{kl}^j] \quad (6)$$

$$D_{pqr}^j = \frac{1}{6}\sum_{P=1}^{N}\sum_{Q=1}^{N}\sum_{R=1}^{N} P_{PQR}^j V_p^P V_q^Q V_r^R \quad (7)$$

And $V_p^P$ is the $p^{th}$ element of the $P^{th}$ left eigenvector and so forth. The following proposition contains the main concept of modal series method.

**Proposition 2.1.** Let the solution of (5) an arbitrary initial conditions such as $Y_0 = VX_0 = U^{-1}X_0 = [y_{1,0}, y_{2,0},..., y_{j,0},..., y_{N,0}]^T$, can be written as follows:

$$y_j(t) = \sum_{m=1}^{\infty} y_j^{(m)}(t) \quad \text{for} \quad j = 1,2,...,n \quad (8)$$

in which $y_j^{(m)}(t)$, contains the terms that depends on any $k$-states multiples of initial conditions. For example for $m = 2$, $y_j^{(2)}(t)$ contains the terms depend on any combination such as $(y_{k,0}, y_{l,0})$ for $k,l = 1,2,...,N$. Now by substituting (8) into (5) we have:

$$\sum_{m=1}^{\infty} \dot{y}_j^{(m)}(t) = \lambda_j \left(\sum_{m=1}^{\infty} y_j^{(m)}(t)\right) + \sum_{k=1}^{N}\sum_{l=1}^{N} C_{kl}^j \left(\left(\sum_{m=1}^{\infty} y_k^{(m)}(t)\right)\left(\sum_{m=1}^{\infty} y_l^{(m)}(t)\right)\right)$$
$$+ \sum_{p=1}^{N}\sum_{q=1}^{N}\sum_{r=1}^{N} D_{pqr}^j \left(\left(\sum_{m=1}^{\infty} y_p^{(m)}(t)\right)\left(\sum_{m=1}^{\infty} y_q^{(m)}(t)\right)\left(\sum_{m=1}^{\infty} y_r^{(m)}(t)\right)\right) + ... \quad (9)$$

then the solution of (9)(and equivalently (5)), can be found by solving the following sequence of differential equations:

$$\dot{y}_j^{(1)} = \lambda_j y_j^{(1)} \tag{10a}$$

$$\dot{y}_j^{(2)} = \lambda_j y_j^{(2)} + \sum_{k=1}^{N}\sum_{l=1}^{N} C_{kl}^j \left( y_k^{(1)} y_l^{(1)} \right) \tag{10b}$$

$$\dot{y}_j^{(3)} = \lambda_j y_j^{(3)} + \sum_{k=1}^{N}\sum_{l=1}^{N} C_{kl}^j \left( y_k^{(1)} y_l^{(2)} + y_l^{(1)} y_k^{(2)} \right) + \sum_{p=1}^{N}\sum_{q=1}^{N}\sum_{r=1}^{N} D_{pqr}^j \left( y_p^{(1)} y_q^{(1)} y_r^{(1)} \right) \tag{10c}$$

$$\vdots$$

with the initial conditions:

$$\begin{cases} y_j^{(1)}(0) = y_{j,0} \\ y_j^{(m)}(0) = 0 \quad, \quad m > 1 \quad, \quad j = 1,2,\dots,N \end{cases} \tag{11}$$

**Proof.** The solution of (5) for some interval $t \in [0,T] \subset R$ for arbitrary initial conditions $Y_0 = [y_{1,0}, y_{2,0}, \dots, y_{j,0}, \dots, y_{N,0}]^T \in \vartheta$ for $j = 1, 2, \dots, N$ can be expressed as:

$$y_j(t) = \Lambda_j(Y_0, t) \tag{12}$$

Where $\Lambda_j : C^N \times R \to C$, is a smooth analytic function with respect to $Y_0$ and $t$. Now we can expand (12) as Maclaurin series with respect to $Y_0$ which yields;

$$y_j(Y_0,t) = \underbrace{\sum_{i=1}^{N} \alpha_{ji} y_{i,0}}_{y_j^{(1)}(t)} + \underbrace{\frac{1}{2}\sum_{k=1}^{N}\sum_{l=1}^{N} \beta_{kl}^j y_{k,0} y_{l,0}}_{y_j^{(2)}(t)} + \underbrace{\frac{1}{6}\sum_{p=1}^{N}\sum_{q=1}^{N}\sum_{r=1}^{N} \gamma_{pqr}^j y_{p,0} y_{q,0} y_{r,0}}_{y_j^{(3)}(t)} + \dots \tag{13}$$

Where, $\alpha_{ji}(t) = \left.\dfrac{\partial \Lambda_j}{\partial y_{i,0}}\right|_{Y_0=0}$, $\beta_{kl}^j(t) = \left.\dfrac{\partial^2 \Lambda_j}{\partial y_{k,0} \partial y_{l,0}}\right|_{Y_0=0}$ and $\gamma_{pqr}^j(t) = \left.\dfrac{\partial^3 \Lambda_j}{\partial y_{p,0} \partial y_{q,0} \partial y_{r,0}}\right|_{Y_0=0}$. Since $\Lambda_j$ is analytic function, existence and uniformly convergence of Maclaurin series in (13) is guaranteed. Now let the initial condition be $\varepsilon Y_0 = [\varepsilon y_{1,0}, \dots, \varepsilon y_{N,0}]^T$ where $\varepsilon$ is an arbitrary scalar parameter. This parameter only simplifies the calculations and its value doesn't have any significant. One can similarly to (13) write:

$$y_j(\varepsilon Y_0, t) = \varepsilon y_j^{(1)}(t) + \varepsilon^2 y_j^{(2)}(t) + \varepsilon^3 y_j^{(3)}(t) + \dots \tag{14}$$

Since $y_j(\varepsilon Y_0, 0) = \varepsilon y_{j,0}$, it follows:

$$\varepsilon y_{j,0} = \varepsilon y_j^{(1)}(0) + \varepsilon^2 y_j^{(2)}(0) + \varepsilon^3 y_j^{(3)}(0) + \dots \tag{15}$$

Substituting (14) into (5) and rearranging with respect to the order of $\varepsilon$ yields:

$$\varepsilon \dot{y}_j^{(1)}(t) + \varepsilon^2 \dot{y}_j^{(2)}(t) + \varepsilon^3 \dot{y}_j^{(3)}(t) + \ldots = \varepsilon\left(\lambda_j y_j^{(1)}\right) + \varepsilon^2\left(\lambda_j y_j^{(2)} + \sum_{k=1}^N \sum_{l=1}^N C_{kl}^j y_k^{(1)} y_l^{(1)}\right) + \\ \varepsilon^3\left(\lambda_j y_j^{(3)} + \sum_{k=1}^N \sum_{l=1}^N C_{kl}^j \left(y_k^{(1)} y_l^{(2)} + y_k^{(2)} y_l^{(1)}\right) + \sum_{p=1}^N \sum_{q=1}^N \sum_{r=1}^N D_{pqr}^j y_p^{(1)} y_q^{(1)} y_r^{(1)}\right) + \ldots \quad (16)$$

Since (16) must be hold for any $\varepsilon$, terms with the same order of $\varepsilon$ on each side must be equal and the proposition is thus proved.

**Remark 2.1.** It should be noted that (10a) yields linear approximate solution to the system, (10b) yields correction terms to linear approximate solution by considering second order nonlinearity, (10c) considers the third order nonlinearity and so on. It is obvious that (10) is a sequence of inhomogeneous linear time invariant ODE's in which at each step, inhomogeneous terms are calculated from the previous step, so the above process is a recursive process.

## 3. APPROXIMATE SOLUTION

In fact, obtaining the exact solution of (5) or equivalently system (3), as in (8) is impossible since (8) contains infinite series. Therefore in practical application by intercepting $M$ terms of the series an approximate closed-form solution can be achieved as follows:

$$y_j(t) \cong [y_j]^{(M)} = \sum_{m=1}^M y_j^{(m)}(t) \quad \text{for} \quad j = 1,2,\ldots,n \quad (12)$$

The integer $M$ is generally determined according to a desirable precision of the approximate solution. In order to obtain an accurate enough approximate closed-from solution, we present an iterative algorithm with low computational complexity as follows:

**Algorithm:**

**Step1.** Let $m = 1$.

**Step2.** Calculate the $m^{th}$ order term $y_j^{(m)}(t)$ from the presented linear ODEs (10) with initial conditions (11) for $j = 1,2,\ldots,N$.

**Step3.** Let $M = i$ and calculate $M^{th}$ order approximate closed-form solution according to (12).

**Step4.** If there was no significant difference between $[y_j]^{(M)}$ and $[y_j]^{(M-1)}$ go to step5; else replace $i$ by $i + 1$ and go to step2.

**Step5.** Stop the algorithm; $[y_j]^{(M)}$ is the desirable approximate solution.

## 4. SIMULATION

By using the above algorithm we found very accurate approximate solution of the Genesio chaotic system just with 3 iterations of the proposed method. In Fig.1, Fig.2 and Fig. 3, it can be observed the validity of the Modal series by comparing to the numerical output of ode45 Matlab software code ode45. Numerical Results show the more accuracy of Modal series in comparison with VIM and MVIM [19] and in comparison with numerical output from ode45. As the time progresses, the accuracy of solution of three dependent variables, $x(t)$, $y(t)$ and $z(t)$ via Modal series is much more than VIM and MVIM. Just 2-iterations of Modal series is enough to obtain an approximate accurate solution in comparison with 5-iterations of VIM and MVIM. However we should mention that on a longer time frame, all the methods miss their accuracy more and more and it can be due to one major issue, which is the chaotic nature of system.

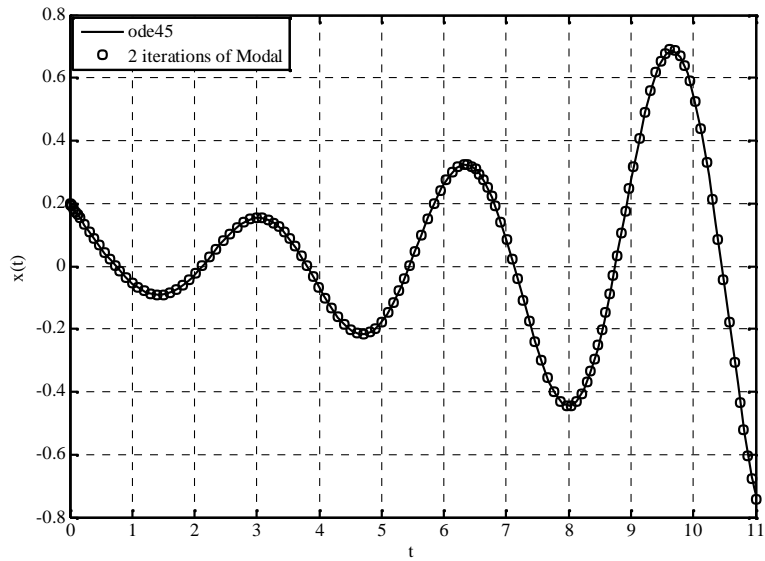

Fig. 1 Comparison of 2-iterations Modal series with ode45 for x(t)

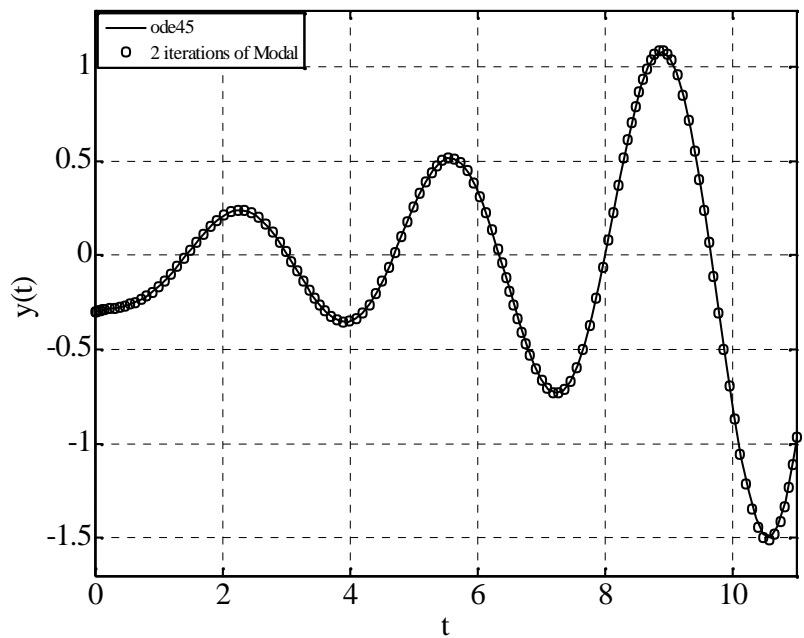

Fig. 2 Comparison of 2-iterations of Modal series with ode45 for y(t)

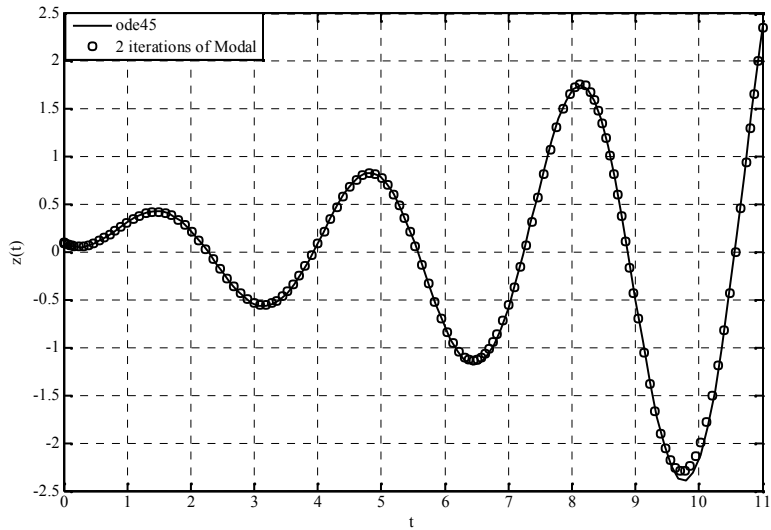

Fig. 3 Comparison of 2-iterations of Modal series with ode45 for z(t)

Table. 1 Comparison of Numerical Results for x(t)

| t | VIM | MVIM | Modal | ode45 | $E_{VIM}$ | $E_{MVIM}$ | $E_{Modal}$ |
|---|---|---|---|---|---|---|---|
| 1 | -0.053918 | -0.054004 | -0.054006 | -0.054004 | 0.000086 | 0 | 0.000002 |
| 2 | -0.029462 | -0.026877 | -0.026835 | -0.026876 | 0.002586 | 0.000001 | 0.000041 |
| 3 | -0.144638 | 0.154580 | 0.154645 | 0.154558 | 0.299196 | 0.000022 | 0.000087 |
| 4 | -2.378007 | 0.067627 | -0.067731 | -0.067725 | 2.310282 | 0.135352 | 0.000007 |
| 5 | -0.586508 | -0.177458 | -0.177253 | -0.177425 | 0.409083 | 0.000033 | 0.000172 |
| 6 | 0.069E03 | 0.260882 | 0.261579 | 0.260981 | 68.73902 | 0.000099 | 0.000598 |
| 7 | 0.479E03 | 0.088230 | 0.086703 | 0.087005 | 478.9130 | 0.001225 | 0.000303 |
| 8 | 0.208E04 | -0.446444 | -0.446720 | -0.446561 | 2080.447 | 0.000117 | 0.000159 |
| 9 | 0.706E04 | 0.272219 | 0.274699 | 0.269644 | 7059.730 | 0.002575 | 0.005055 |
| 10 | 0.203E05 | 0.551331 | 0.548076 | 0.548875 | 20299.45 | 0.002456 | 0.000799 |
| 11 | 0.519E05 | -0.766691 | -0.743769 | -0.732707 | 51900.73 | 0.033984 | 0.011063 |

$E_{VIM} = |x_{VIM} - x_{ode45}|$    $E_{MVIM} = |x_{MVIM} - x_{ode45}|$    $E_{Modal} = |x_{Modal} - x_{ode45}|$

Table. 2 Comparison of Numerical Results for y(t)

| t | VIM | MVIM | Modal | ode45 | $E_{VIM}$ | $E_{MVIM}$ | $E_{Modal}$ |
|---|---|---|---|---|---|---|---|
| 1 | -0.165953 | -0.166059 | -0.166063 | -0.166059 | 0.000106 | 0 | 0.000002 |
| 2 | 0.219206 | 0.208674 | 0.208781 | 0.208674 | 0.010532 | 0 | 0.000107 |
| 3 | 0.319313 | 0.016564 | 0.016464 | 0.016537 | 0.302776 | 0.000027 | 0.000073 |
| 4 | 3.694111 | 0.016564 | -0.349846 | -0.349762 | 4.043873 | 0.366326 | 0.000084 |
| 5 | 31.523180 | 0.254818 | 0.255746 | 0.255264 | 31.26792 | 0.000446 | 0.000482 |
| 6 | 0.228E03 | 0.344919 | 0.343750 | 0.343848 | 227.6562 | 0.001071 | 0.000098 |
| 7 | 0.157E04 | -0.658754 | -0.660823 | -0.659586 | 1570.660 | 0.000832 | 0.001237 |
| 8 | 0.881E04 | -0.004417 | 0.000915 | -0.001932 | 8810.002 | 0.002485 | 0.002848 |
| 9 | 0.394E05 | 1.053578 | 1.057037 | 1.053196 | 39398.95 | 0.000382 | 0.003842 |
| 10 | 0.146E06 | -0.774085 | -0.812579 | -0.797749 | 146000.8 | 0.023664 | 0.014829 |
| 11 | 0.464E06 | -1.261695 | -0.963112 | -0.970215 | 464001.0 | 0.291480 | 0.007103 |

$E_{VIM} = |y_{VIM} - y_{ode45}|$    $E_{MVIM} = |y_{MVIM} - y_{ode45}|$    $E_{Modal} = |y_{Modal} - y_{ode45}|$

Table. 3 Comparison of Numerical Results for z(t)

| t | VIM | MVIM | Modal | ode45 | $E_{VIM}$ | $E_{MVIM}$ | $E_{Modal}$ |
|---|---|---|---|---|---|---|---|
| 1 | 0.311946 | 0.312200 | 0.312219 | 0.312200 | 0.000254 | 0 | 0.000019 |
| 2 | 0.224912 | 0.229696 | 0.229758 | 0.229692 | 0.004780 | 0.000004 | 0.000066 |
| 3 | -0.355134 | -0.542526 | -0.542762 | -0.542542 | 0.187408 | 0.000016 | 0.000220 |
| 4 | 2.716517 | 0.097729 | 0.098160 | 0.098060 | 2.618457 | 0.000331 | 0.000010 |
| 5 | -12.03379 | 0.780386 | 0.781172 | 0.780235 | 12.81403 | 0.000151 | 0.000936 |
| 6 | 0.428E03 | -0.781967 | -0.785908 | -0.783868 | 428.7839 | 0.001901 | 0.002040 |
| 7 | 0.465E05 | -0.580459 | -0.576883 | -0.577194 | 46500.58 | 0.003265 | 0.000311 |
| 8 | 0.107E07 | 1.697533 | 1.701202 | 1.692868 | 1069998 | 0.004665 | 0.008334 |
| 9 | 0.139E08 | -0.508330 | -0.489641 | -0.476252 | 1390000 | 0.032078 | 0.013389 |
| 10 | 0.126E09 | -2.094007 | -2.175953 | -2.165551 | 12600000 | 0.071544 | 0.010402 |
| 11 | 0.886E09 | 1.247071 | 2.446294 | 2.378996 | 88599999 | 1.131926 | 0.067298 |

$$E_{VIM} = |z_{VIM} - z_{ode45}| \qquad E_{MVIM} = |z_{Modal} - z_{ode45}| \qquad E_{Modal} = |z_{Modal} - z_{ode45}|$$

As it can be seen, the precision of 2-iterations of Modal is close to 5-ietarations of MVIM in comparison with ode45. As the time progress VIM is completely invalid whereas Modal and MVIM agree well to the numerical solution.

## 5. CONCLUSIONS

The main objective of this paper was to show the efficiency of the Modal series method for solving the nonlinear chaotic systems. By using the Modal series method the exact solution of nonlinear chaotic systems can be determined in the form of uniformly convergent series. Besides for obtaining an approximate solution with easy computable terms, a straightforward algorithm with low computational complexity was brought. The proposed method only requires solving a sequence of linear time-invariant ODEs in a recursive manner. Thus in comparison to other approximate methods, this method is more accurate and more practical. Also uniformly convergent of this method is guaranteed unlike other famous methods such as HPM or VIM. Numerical simulations are evidences to this assertion.

## REFERENCES

[1] S.M. Roberts, J.S. Shipman, Two Point Boundary Value Problems: Shooting Methods, American Elsevier, New York, 1972.
[2] Morrison, D. D., Riley, J. D., and Zancanaro, J. F., "Multiple Shooting Method for Two-Point Boundary Value Problems," Communications of the ACM, 1962, pp. 613– 614.
[3] Holt, J. F., "Numerical Solution of Nonlinear Two-Point Boundary Problems by Finite Difference Methods," Communications of the ACM, 1964, pp. 366–373.
[4] G. Adomian, Solving Frontier Problems of Physics: The Decomposition Method, Kluwer Academic, Dordrecht, 1994.
[5] G. Adomian, R. Rach, Modified decomposition solution of linear and nonlinear boundary-value problems, Nonlinear Anal. 23 (5) (1994) 615–619.
[6] Hashim, I., M.S.M. Noorani, R. Ahmad, S.A. Bakar, E.S.I. Ismail and A.M. Zakaria, 2006. Accuracy of the Adomian decomposition method applied to the Lorenz system. Chaos Soliton Fract., 28: 1149-1158. DOI: 10.1016/j.chaos.2005.08.135
[7] 6. Batiha, B., M.S.M. Noorani and I. Hashim, 2007. Numerical solution of sine Gordon equation by variational iteration method. Phys. Lett. A., 370: 437-440. DOI: org/10.1016/j.physleta. 2007.05.087
[8] S. Momani, S. Abuasad, Z. Odibat, Variational iteration method for solving nonlinear boundary value problems, Appl. Math. Comput. 183 (2006) 1351– 1358.
[9] J. Lu, Variational iteration method for solving two-point boundary value problems, J. Comput. Appl. Math. 207 (2007) 92–95.

[10] 11. Alomari, A.K., M.S.M. Noorani and R. Nazar, 2008. Adaptation of homotopy analysis method for the numeric-analytic solution of Chen system. Commun. Nonlinear Sci. Numer. Simul., 14: 1346-1354. DOI: 10.1016/j.cnsns. 2008.02.007
[11] S.J. Liao, Beyond Perturbation: Introduction to the Homotopy Analysis Method, Chapman & Hall/CRC Press, Boca Raton, 2003.
[12] A.H. Nayfeh, Perturbation Methods, Wiley, New York, 2000.
[13] J.H. He, Homotopy perturbation method for solving boundary value problems, Phys. Lett. A350 (1–2) (2006) 87–88.
[14] J.H. He, Homotopy perturbation method: a new nonlinear analytical technique, Appl. Math. Comput. 135 (2003) 73–79.
[15] 9. Chowdhury, M.S.H. and I. Hashim, 2007. Application of multistage homotopy-perturbation method for the solutions of the Chen system. Nonlinear Anal.: Real World Appli., 10: 381-391.DOI: 10.1016/j.nonrwa.2007.09.014
[16] N. Pariz, H. M. Shanechi and E. Vaahedi, Explaining and validating stressed power systems behavior using modal series, IEEE T. Power Syst., vol.18, no.2, pp.778–785, 2003.
[17] H. M. Shanechi, N. Pariz and E. Vahedi, General nonlinear modal representation of large scale power systems, IEEE T. Power Syst., vol.18, no.3, pp.1103-1109, 2003.
[18] F. X. Wu, H. Wu, Z. X. Han and D. Q. Gan, Validation of power system non-linear modal analysis methods, Electr. Pow. Syst. Res., vol.77, no.10, pp.1418-1424, 2007.